\documentclass[pra,twocolumn,superscriptaddress,preprintnumbers,amsmath,amssymb]{revtex4}
\usepackage{}

\usepackage{amsfonts}
\usepackage{amsmath}
\usepackage{amssymb}
\usepackage{amstext}
\usepackage{amsthm}
\usepackage{bm}% bold math
\usepackage{bbm}
\usepackage{booktabs}
\usepackage{braket}
\usepackage{color}%
\usepackage{dcolumn}
\usepackage{dsfont}
\usepackage{epsfig,graphicx}
\usepackage{float}
\usepackage{graphicx}
\usepackage{mathrsfs}
\usepackage{mathtools}
\usepackage{mleftright}
\usepackage{times}
\usepackage{txfonts}
\usepackage[utf8]{inputenc}

\usepackage{hyperref}
\hypersetup{colorlinks=true,linkcolor=blue,citecolor=blue,urlcolor=blue}

\setcitestyle{open={[},close={]},citesep={,\!},numbers}

\AtBeginDocument{%
\setlength{\abovedisplayskip}{4pt plus 1pt minus 1pt}%
\setlength{\belowdisplayskip}{4pt plus 1pt minus 1pt}%
\setlength{\abovedisplayshortskip}{2pt plus 1pt}%
\setlength{\belowdisplayshortskip}{2pt plus 1pt}%
\setlength{\textfloatsep}{10pt plus 2pt minus 2pt}%
\setlength{\intextsep}{10pt plus 2pt minus 2pt}%
\setlength{\floatsep}{10pt plus 2pt minus 2pt}%
\setlength{\abovecaptionskip}{6pt plus 1pt minus 1pt}%
\setlength{\belowcaptionskip}{0pt}%
\typeout{===JJC check: abovedisplayskip=\the\abovedisplayskip\space belowdisplayskip=\the\belowdisplayskip\space textfloatsep=\the\textfloatsep===}%
}

\begin{document}

\title{Bound-state-mediated remote charging of a quantum battery}

\author{Jian-Jian Cheng}
\affiliation{School of Science, Xi'an University of Posts and Telecommunications, Xi'an 710121, China}

\author{Hai-Bo Qiu}
\affiliation{School of Science, Xi'an University of Posts and Telecommunications, Xi'an 710121, China}

\author{Lin Zhang}
\affiliation{School of Physics and Information Technology, Shaanxi Normal University, Xi'an 710119, China}

\author{Ming-Liang Hu}
\email{mingliang0301@163.com}
\affiliation{School of Science, Xi'an University of Posts and Telecommunications, Xi'an 710121, China}

\begin{abstract}
Remote charging of a quantum battery (QB) is hindered by radiative leakage of the excitation into the photonic environment that acts as a mediator for energy transfer. We consider a charger-battery model consisting of two two-level systems (TLSs) that are locally coupled to two sites of a one-dimensional coupled cavity array. When their transition frequency lies outside the propagation band, the system forms atom-photon bound states with localized photonic components, and the overlap of these components lifts the degeneracy of the even- and odd-parity bound states, yielding an energy splitting that drives coherent energy transfer from the charger to the QB. In this way, the band gap suppresses resonant emission and the localized bound states mediate remote charging. From the parity-resolved spectrum, we relate the charging time to the energy splitting and the charged ergotropy to the fraction of TLS population on the bound states. Bound states closer to the band edge extend the interaction range of the TLSs but contain a large photonic fraction and are consequently more susceptible to photon loss.
\end{abstract}

\maketitle

\section{Introduction}
%%%%%%%%%%%%%%%%%%%%%%%%%%%%%%%%%%%%%%%%%%%%%%%%%%%%%%%%%%%%%%%%%%%%%
Quantum battery (QB) stores and releases energy in a controllable microscopic system. The extractable work of the QB, called ergotropy, is defined as the maximum amount of energy extractable via a unitary operation \cite{Alicki2013, Allahverdyan2004}. Quantum coherence and collective effects can enhance the charging power and, in suitable settings, the extractable work \cite{Ferraro2018,Campaioli2017,Andolina2018,Andolina2019, Gyhm2022}. Physically, QB can be implemented in few-level systems, spin systems, oscillators, and cavity-QED architectures \cite{Binder2015,Campaioli2024,Hovhannisyan2013,Le2018,Crescente2020,Dou2022a,Dou2022b,Liu2021, Rossini2019,Hu2025b}. The experimental realizations of a QB have also been reported in the NMR, superconducting, and photonic platforms \cite{Joshi2022,Quach2022,Ferraro2022,Xue2023}.

In realistic settings, the energy relaxation and decoherence may result in aging of the QB \cite{Pirmoradian2019}. The schemes such as dark states, decoherence-free subspace, Floquet control, reservoir engineering, and feedback control have been explored to mitigate these adverse effects \cite{Quach2020,Bai2020,Xu2021, Mitchison2021,Barra2019,Farina2019,GarciaPintos2020}. For the remote charging, the difficulty is more apparent: the photonic environment connecting the charger and QB can also carry the excitation away from the local storage degrees of freedom. When the common transition frequency of the charger and the QB lies inside the propagation band, the excitation overlaps with the extended lattice modes and thereby becomes exposed to propagation loss and retardation, the effect that becomes increasingly relevant as the separation between the charger and the QB grows.

The structured photonic environments \cite{Chang2018,Sheremet2023,Lodahl2015} offer several pathways to reduce the radiative loss. In particular, the remote charging of a QB has recently been studied \cite{Song2024}. More broadly, the out-of-band bound states can yield long-lived charger-battery-field correlations \cite{John1990,Liu2017,Calajo2016,Shi2016,Leonforte2021,GonzalezTudela2017}. Additionally, the topological structures provide edge or dark states robust against certain classes of disorder \cite{Lu2025,Bello2019,Su1979}, and the giant atoms can suppress radiative loss by interference among the coupling points \cite{Yan2026,Kockum2018,Kockum2014}. The wireless charging of a QB via the common bosonic reservoir has also been investigated \cite{Hu2025a}.

In this study, we consider two spatially separated two-level systems (TLSs), which act as the charger and QB and are locally coupled to a one-dimensional coupled cavity array. Outside the propagation band, this system supports atom-photon bound states with localized photonic components. The spatial overlap of these components leads to a distance-dependent effective interaction of the TLSs and lifts the degeneracy of the even- and odd-parity bound states, and the resulting beating converts this static overlap into dynamical energy transfer. Because this component has a finite occupation, it remains subject to cavity photon loss. We will derive the parity-resolved spectrum, connect it to the charged ergotropy of the QB, and determine how the TLS relaxation and cavity photon loss limit the remote charging performance.

\section{The charger-battery model and battery performance}
\begin{figure}[!htbp]
\centering
\includegraphics[width=\columnwidth]{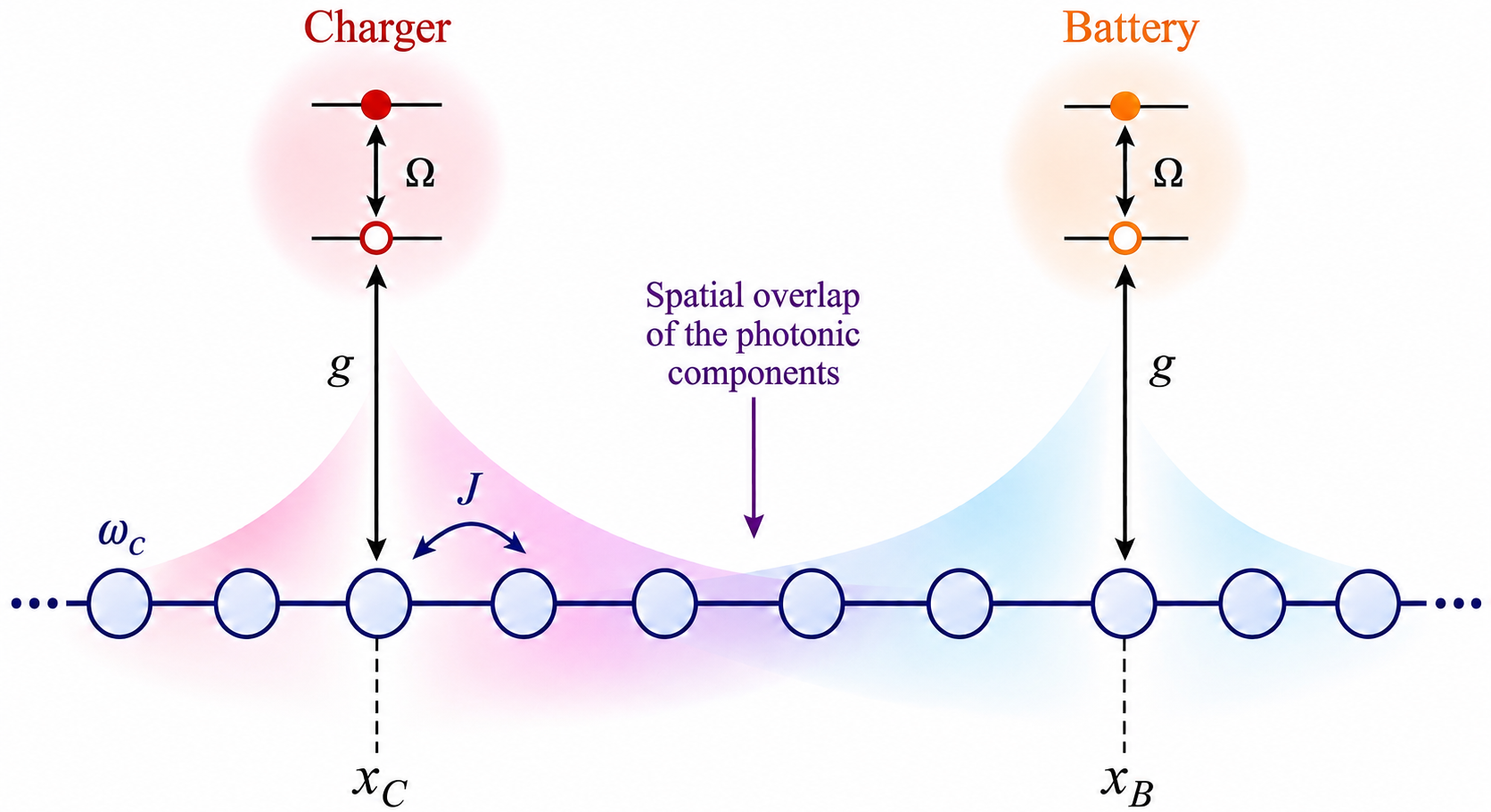}
\caption{Schematic of remote charging, where the charger and battery are two spatially separated TLSs without direct coupling. They have transition frequency $\Omega$ and are locally coupled to a one-dimensional coupled-cavity array at sites $x_C$ and $x_B$, with the coupling strength $g$ and separation $d=|x_B-x_C|$. Each cavity has frequency $\omega_c$, and the hopping amplitude between the neighboring cavities is $J$.} \label{fig:mechanism}
\end{figure}

%%%%%%%%%%%%%%%%%%%%%%%%%%%%%%%%%%%%%%%%%%%%%%%%%%%%%%%%%%%%%%%%%%%%%
As sketched in Fig. \ref{fig:mechanism}, we model the charger $(C)$ and battery $(B)$ as two TLSs without direct exchange coupling. The TLSs are locally coupled to a one-dimensional coupled cavity array at sites $x_C$ and $x_B$, respectively. The superconducting coupled resonator array provides a direct implementation of the tight-binding Hamiltonian below \cite{Mirhosseini2018,Anderson2016}, with the transmon or flux qubits acting as the TLSs \cite{Blais2004,Devoret2013}, and the related dispersions can be engineered in photonic crystal waveguides \cite{Liu2017}.

% For one-column wide figures use

In the rotating wave approximation, the total Hamiltonian (in units of $\hbar$) in the real space can be written as
%%%%%%%%%%%%%%%%%%%%%%%%%%%
\begin{equation} \label{eq:hamiltonian}
\begin{aligned}
 H = & \; \omega_c \sum_x a_x^\dagger a_x - J \sum_x (a_x^\dagger a_{x+1} + a_x a_{x+1}^\dagger) \\
     & + \Omega (\sigma_C^{+} \sigma_C^{-} + \sigma_B^{+} \sigma_B^{-})
       + g\sum_{\nu=C,B} (a_{x_\nu}\sigma_\nu^{+} + a_{x_\nu}^\dagger\sigma_\nu^{-}),
\end{aligned}
\end{equation}
%%%%%%%%%%%%%%%%%%%%%%%%%%%
where $a_x$ ($a_x^\dagger$) is the annihilation (creation) operator of the $x$th cavity mode with frequency $\omega_c$, $J$ is the nearest-neighbor hopping amplitude, $\Omega$ is the transition frequency of the two TLSs, with the raising (lowering) operator $\sigma_\nu^{+}$ ($\sigma_\nu^{-}$) ($\nu=C,B$), and $g$ is their coupling strength to the cavity.

For the bare cavity array, we set the lattice spacing to unity and take $k\in[-\pi,\pi)$ in the first Brillouin zone. The Fourier transformation $a_k = N^{-1/2}\sum_x e^{ikx}a_x$ (with $N$ being the number of cavities) diagonalizes the bare cavity array Hamiltonian and gives the tight-binding dispersion $\omega_k = \omega_c-2J\cos k$ \cite{Zhou2008}, then in the frame rotating at $\omega_c$, the photonic continuum occupies only a finite band $E\in[-2J,2J]$. This finite bandwidth is the basis of band-gap charging: by suitably choosing the TLS frequency and coupling, the associate dressed energies can be placed outside the continuum (i.e., $|E|>2J$), and in this out-of-band region, the resonant emission into propagating modes is suppressed by the band structure \cite{John1990,Liu2017}. The same finite bandwidth also allows the localized dressed eigenstates to form outside the continuum, which provides the charging channel analyzed below. When the TLS transition lies inside the propagation band, however, the initial excitation overlaps with the extended eigenmodes and may decay away from the local charger-battery subspace.

We assess the charging performance via the energy and ergotropy. The mean energy of a QB is given by
%%%%%%%%%%%%%%%%%%%%%%%%%%%
\begin{equation} \label{eq:EB-def}
E_B(t)= \mathrm{Tr} [\rho_B(t)H_B],
\end{equation}
%%%%%%%%%%%%%%%%%%%%%%%%%%%
where $H_B= \Omega \sigma_B^+\sigma_B^-$ is the battery Hamiltonian and $\rho_B(t)$ is the battery state obtained by tracing out the charger and cavity array. In general, not all the stored energy can be extracted as work. The ergotropy quantifies the maximal amount of work extractable from a QB via a unitary transformation, which is given by \cite{Allahverdyan2004,Alicki2013,Francica2020,Cakmak2020, Shi2022}
%%%%%%%%%%%%%%%%%%%%%%%%%%%
\begin{equation} \label{eq:WB-def}
 \mathcal{W}_B(t) = \mathrm{Tr} [\rho_B(t)H_B] - \mathrm{Tr} [\tilde{\rho}_B(t)H_B],
\end{equation}
%%%%%%%%%%%%%%%%%%%%%%%%%%%
with $\tilde{\rho}_B(t)=\sum_n r_n(t) |\varepsilon_n\rangle \langle\varepsilon_n|$ the passive state associated with $\rho_B(t)$. Here, $r_n(t)$ are the eigenvalues of $\rho_B(t)$ arranged in nonincreasing order, while $|\varepsilon_n\rangle$ are the eigenstates of $H_B$ with the associated eigenvalues $\varepsilon_n$ arranged in nondecreasing order.

\section{Bound-state structure and remote charging}
%%%%%%%%%%%%%%%%%%%%%%%%%%%%%%%%%%%%%%%%%%%%%%%%%%%%%%%%%%%%%%%%%%%%%
We take the initial system state to be
%%%%%%%%%%%%%%%%%%%%%%%%%%%
\begin{equation} \label{eq:init-state}
 |\psi(0)\rangle= |e_C, g_B, 0_\mathrm{lat}\rangle,
\end{equation}
%%%%%%%%%%%%%%%%%%%%%%%%%%%
which describes an excited charger, an empty QB, and the cavity array is in the vacuum. For this case, the system dynamics is confined to the single-excitation sector, within which the exchange symmetry of the two TLSs divides the problem into the even- and odd-parity cases \cite{Zhou2009,Zhou2008b,Liao2010}. In the frame rotating at the cavity frequency $\omega_c$, the Hamiltonian becomes
%%%%%%%%%%%%%%%%%%%%%%%%%%%
\begin{equation} \label{eq:H-k}
\begin{aligned}
 H_\mathrm{rot} = {}& -2J\sum_k \cos(k) a_k^\dagger a_k + \delta (\sigma_C^{+} \sigma_C^{-} + \sigma_B^{+} \sigma_B^{-}) \\
                  {}& + \frac{g}{\sqrt{N}} \sum_{k} \big[a_k (\sigma_C^{+} e^{-ikx_C} + \sigma_B^{+} e^{-ikx_B})+\mathrm{H.c.}\big],
\end{aligned}
\end{equation}
%%%%%%%%%%%%%%%%%%%%%%%%%%%
with $\delta = \Omega-\omega_c$. In this rotating frame the bound-state energies $E_{\pm,s}$ are measured relative to $\omega_c$, whereas the local battery energy $E_B$ is evaluated using the bare $H_B$, with zero energy assigned to the battery ground state. We introduce the even- and odd-parity combinations
%%%%%%%%%%%%%%%%%%%%%%%%%%%
\begin{equation} \label{eq:As-def}
 |A_s\rangle = \frac{1}{\sqrt{2}} (\sigma_C^{+} \pm \sigma_B^{+}) |g_C, g_B, 0_\mathrm{lat}\rangle \hspace{0.6em} (s=e,o),
\end{equation}
%%%%%%%%%%%%%%%%%%%%%%%%%%%
where the `$+$' (`$-$') sign corresponds to the even-parity (odd-parity) state with $s=e$ ($s=o$) throughout this paper. These states are, respectively, the in-phase and out-of-phase superpositions of the two local TLS excitations. The corresponding single-excitation bound states read
%%%%%%%%%%%%%%%%%%%%%%%%%%%
\begin{equation} \label{eq:phis-def}
 |\phi_s\rangle = b_s |A_s\rangle + \sum_k c_{k,s} a_k^\dagger |g_C,g_B,0_\mathrm{lat}\rangle.
\end{equation}
%%%%%%%%%%%%%%%%%%%%%%%%%%%

Substituting Eq. \eqref{eq:phis-def} into the stationary Schr\"{o}dinger equation $H_\mathrm{rot}|\phi_s\rangle=E_s|\phi_s\rangle$ and matching the atomic and photonic coefficients separately gives
%%%%%%%%%%%%%%%%%%%%%%%%%%%
\begin{equation} \label{eq:coeff}
\begin{aligned}
 & (E_s-\delta)b_s = \frac{g}{\sqrt{2N}} \sum_k c_{k,s} (e^{-ikx_C} \pm e^{-ikx_B}), \\
 & (E_s+2J\cos k)c_{k,s} = \frac{gb_s}{\sqrt{2N}} (e^{ikx_C}\pm e^{ikx_B}).
\end{aligned}
\end{equation}
%%%%%%%%%%%%%%%%%%%%%%%%%%%
Solving the second line of Eq. \eqref{eq:coeff} for $c_{k,s}$ and substituting it into the first line yields the self-consistency relation for the energy in each state,
%%%%%%%%%%%%%%%%%%%%%%%%%%%
\begin{equation} \label{eq:self-consistency}
 E_s-\delta = \frac{g^2}{N} \sum_k \frac{1\pm \cos(kd)} {E_s+2J\cos k},
\end{equation}
%%%%%%%%%%%%%%%%%%%%%%%%%%%
where $d=|x_B-x_C|$, and with the same sign convention as in Eq. \eqref{eq:As-def}, the factors $1\pm\cos(kd)$ arise from the coherent addition or subtraction of the two local coupling amplitudes.

For the out-of-band bound states, the photonic component has an exponentially decaying evanescent profile. In the thermodynamic limit, $N^{-1}\sum_k$ becomes $(2\pi)^{-1}\int_{-\pi}^{\pi} dk$, and Eq. \eqref{eq:self-consistency} reduces to the pole equation
%%%%%%%%%%%%%%%%%%%%%%%%%%%
\begin{equation} \label{eq:pole}
 E_s-\delta = g^2 [G_0(E_s)\pm G_d(E_s)],
\end{equation}
%%%%%%%%%%%%%%%%%%%%%%%%%%%
with the local and nonlocal Green functions given by
%%%%%%%%%%%%%%%%%%%%%%%%%%%
\begin{equation} \label{eq:lattice-green-functions}
\begin{aligned}
 & G_0(E)= \frac{1}{2\pi} \int_{-\pi}^{\pi} \frac{dk}{E+2J\cos k}, \\
 & G_d(E)= \frac{1}{2\pi} \int_{-\pi}^{\pi} \frac{\cos(kd)\,dk}{E+2J\cos k}.
\end{aligned}
\end{equation}
%%%%%%%%%%%%%%%%%%%%%%%%%%%
This decomposition separates two physically distinct effects: $G_0(E)$ dresses each TLS independently, whereas $G_d(E)$ induces coupling of the two distant excitations through the overlap of their localized photonic components.

For $|E|>2J$, the denominator $E+2J\cos k$ does not vanish for any real $k$. The local Green function is
%%%%%%%%%%%%%%%%%%%%%%%%%%%
\begin{equation} \label{eq:G0}
 G_0(E) = \frac{1}{E\sqrt{1-4J^2/E^2}}.
\end{equation}
%%%%%%%%%%%%%%%%%%%%%%%%%%%
It is negative below the band and positive above it. The corresponding localization length $\lambda(E)= 1/\operatorname{arccosh}(|E|/2J)$. For the lower-band states, $E<-2J$, the nonlocal Green function
%%%%%%%%%%%%%%%%%%%%%%%%%%%
\begin{equation} \label{eq:Gd-lower}
 G_d(E) = G_0(E)e^{-d/\lambda(E)},
\end{equation}
%%%%%%%%%%%%%%%%%%%%%%%%%%%
while for the upper-band states ($E>2J$), the photonic amplitude alternates in sign between the neighboring cavities, thus $G_d(E)=(-1)^d G_0(E)e^{-d/\lambda(E)}$. The exponential factor $e^{-d/\lambda(E)}$ sets the spatial range of remote coupling. Hence the even- and odd-parity upper-band branches exchange their order for odd $d$. The even-odd splitting in Eq. \eqref{eq:pole} is a spectral signature of this remote coupling, and for weak overlap, it inherits the exponential distance dependence of $G_d(E)$. For $E_{-,s}<-2J$, substituting Eq. \eqref{eq:Gd-lower} into Eq. \eqref{eq:pole} gives
%%%%%%%%%%%%%%%%%%%%%%%%%%%
\begin{equation} \label{eq:E-band}
\begin{gathered}
 E_{-,e} - \delta = \frac{2g^2 e^{-\frac{d}{2\lambda_{-,e}}} \cosh \left(\frac{d}{2\lambda_{-,e}}\right)} {E_{-,e}\sqrt{1-4J^2/E_{-,e}^2}}, \\% [4pt]
 E_{-,o} - \delta = \frac{2g^2 e^{-\frac{d}{2\lambda_{-,o}}} \sinh \left(\frac{d}{2\lambda_{-,o}}\right)} {E_{-,o}\sqrt{1-4J^2/E_{-,o}^2}},
\end{gathered}
\end{equation}
%%%%%%%%%%%%%%%%%%%%%%%%%%%
where $\lambda_{-,s}= 1/\operatorname{arccosh}(|E_{-,s}|/2J)$ ($s=e,o$) is the localization length of the lower-band states.

% For one-column wide figures use
\begin{figure}[!htbp]
\centering
\includegraphics[width=\columnwidth]{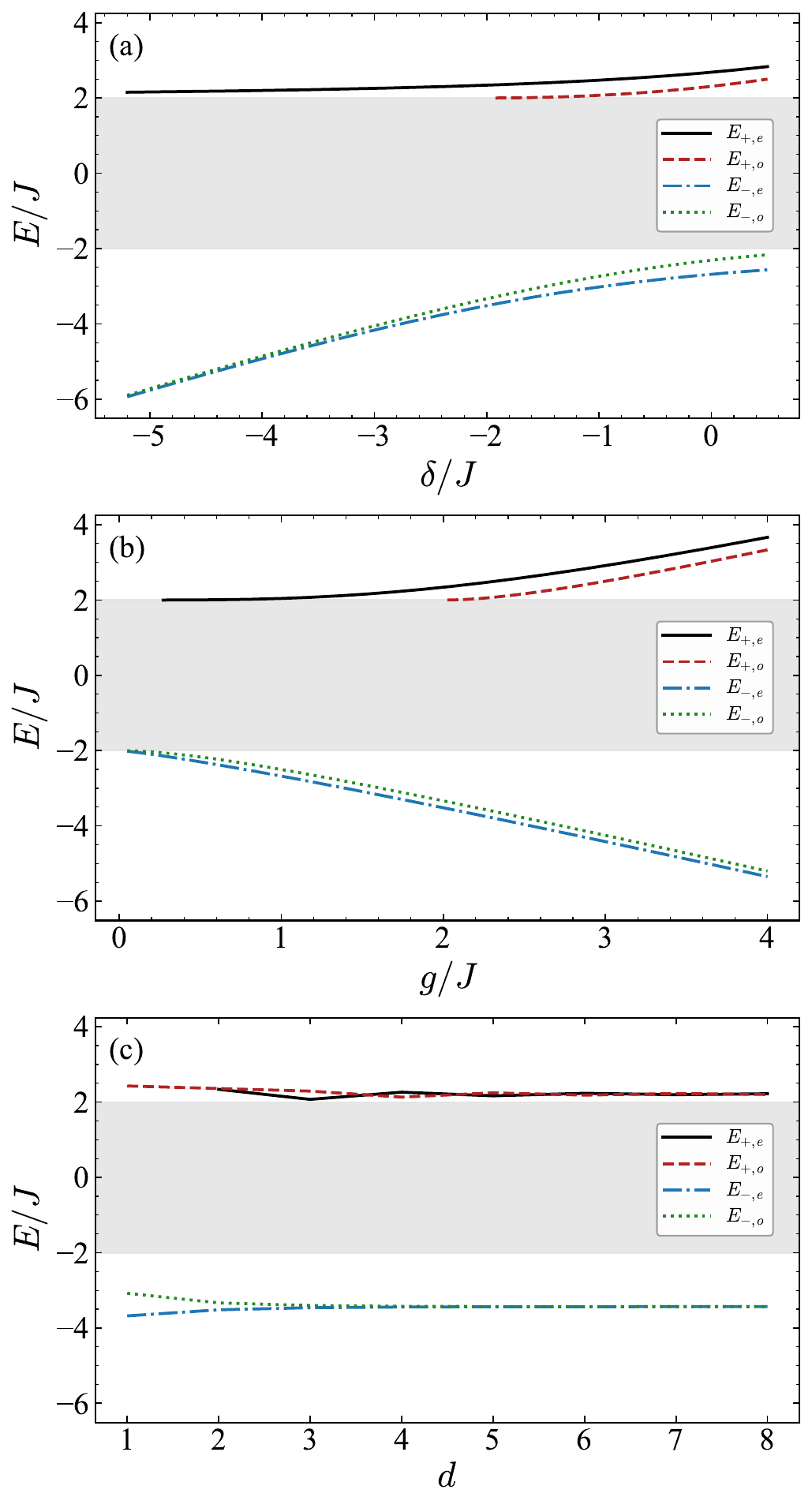}
\caption{Out-of-band bound-state spectrum. The gray shaded region marks the photonic continuum $-2J<E<2J$. The four lines outside it are the upper-band even-parity $E_{+,e}$ and odd-parity $E_{+,o}$, the lower-band even-parity $E_{-,e}$ and odd-parity $E_{-,o}$ branches. (a) $E/J$ vs $\delta/J$ at fixed $g/J=2$ and $d=2$. (b) $E/J$ vs $g/J$ at fixed $\delta/J=-2$ and $d=2$. (c) $E/J$ vs $d$ at fixed $\delta/J=-2$ and $g/J=2$.} \label{fig:bound-spectrum}
\end{figure}

For the parameter ranges considered here, each parity channel supports one bound state above the band, $E_{+,s}>2J$, and one below, $E_{-,s}<-2J$, giving the four branches $E_{+,e}$, $E_{+,o}$, $E_{-,e}$, and $E_{-,o}$. Their parity is assigned by the overlap of the exact eigenstate with $|A_e\rangle$ and $|A_o\rangle$, which automatically tracks the parity exchange of the upper-band roots at odd $d$. Figure \ref{fig:bound-spectrum} shows the four branches obtained by exact diagonalization. In the lower-band-gap regime $\delta/J\lesssim-2$, the two lower-band states remain outside the continuum and retain a finite even-odd separation at short distance. The even and odd combinations are formed by the sum and difference of the two exponential tails, producing the $\cosh$ and $\sinh$ factors in Eq. \eqref{eq:E-band}. Increasing $d$ leaves the individual bound states essentially intact but suppresses their hybridization, thereby the two parity branches approach degeneracy. Remote charging is then hindered due to the collapse of the splitting, rather than due to the  disappearance of the local bound states.

Two limits follow from Eq. \eqref{eq:E-band}. For $d\gg\lambda_{-,s}$, $G_d$ is exponentially small, the even- and odd-parity bound states become nearly degenerate, and coherent energy transfer from the charger to the QB is suppressed. For $d\lesssim\lambda_{-,s}$, the overlap is appreciable, the even-odd splitting is finite, and the charger and QB excitations hybridize into nonlocal dressed states capable of transferring excitation between the two sites. The localization length $\lambda_{-,s}$ is set self-consistently by the bound-state energy rather than by the detuning alone: it depends on $\delta$, $g$, $d$, and the parity, and it controls the spatial separation of remote coupling. For fixed $g$ and $d$, the deeper negative detuning moves the lower bound states far from the band edge, reducing $\lambda_{-,s}$ and increasing the total TLS population.

Choosing the charger site at $x_C=0$ and the battery site at $x_B=d$, then for the lower-band bound states, $E_{-,s}<-2J$, one has
%%%%%%%%%%%%%%%%%%%%%%%%%%%
\begin{equation} \label{eq:phi-real}
\begin{aligned}
 |\phi_{-,s}\rangle = & \; \mathcal{N}_s \Bigg[|A_s\rangle + \frac{g}{E_{-,s}\sqrt{2(1- 4J^2/E_{-,s}^2)}} \\
                      & \times \sum_x \Big(e^{-\frac{|x|}{\lambda_{-,s}}} \pm e^{-\frac{|x-d|}{\lambda_{-,s}}}\Big)
                        a_x^\dagger |g_C,g_B,0_\mathrm{lat}\rangle \Bigg],
\end{aligned}
\end{equation}
%%%%%%%%%%%%%%%%%%%%%%%%%%%
where $\mathcal{N}_s$ normalizes the atom-photon bound state. We denote by $u_x$ the coefficient of $a_x^\dagger |g_C,g_B,0_\mathrm{lat}\rangle$ in Eq. \eqref{eq:phi-real}, so that $|u_x|^2$ is the real-space photon distribution.

% For one-column wide figures use
\begin{figure}[!htbp]
\centering
\includegraphics[width=\columnwidth]{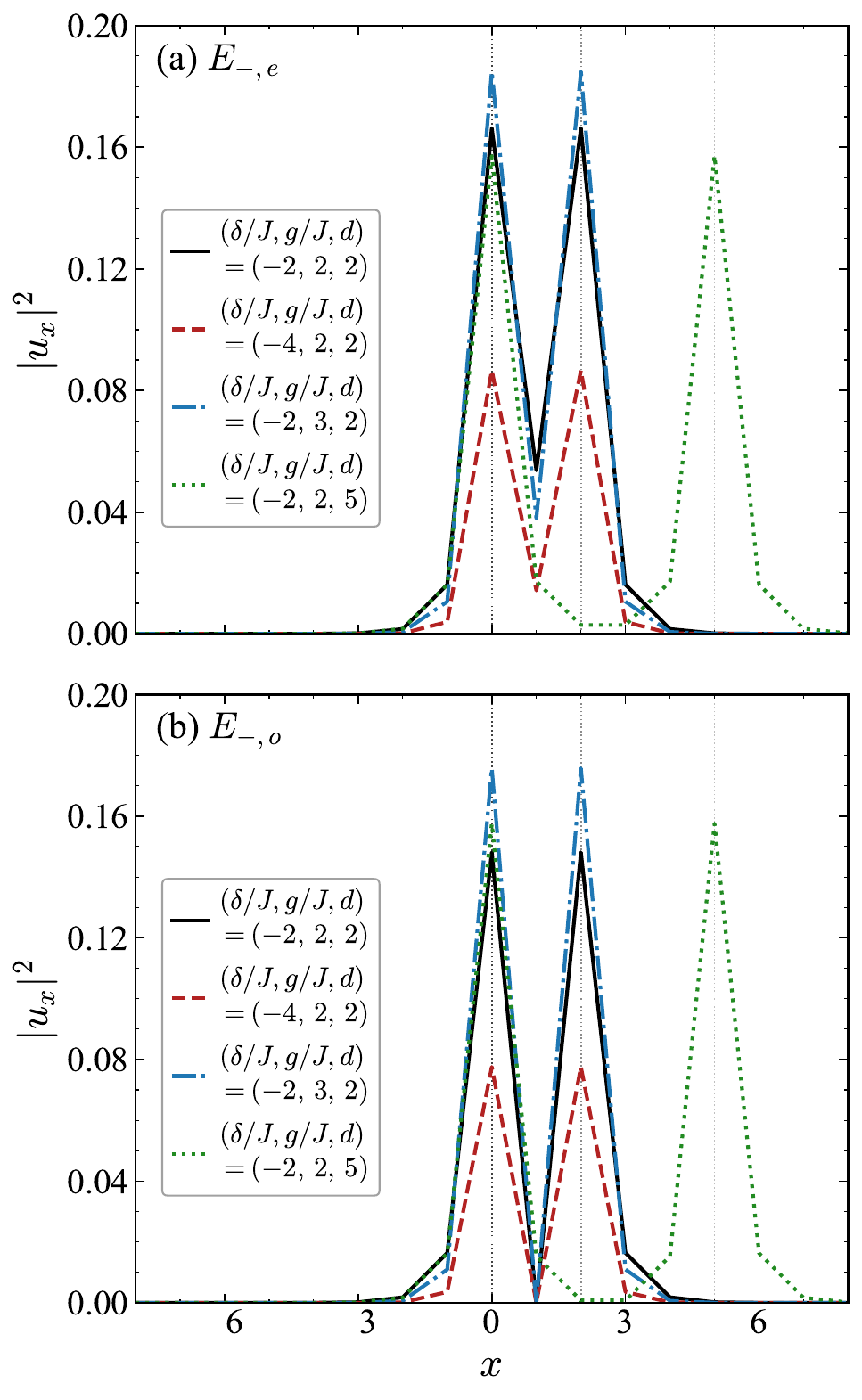}
\caption{Real-space photon distribution $|u_x|^2$ of the lower-band even-parity bound state $E_{-,e}$ (a) and odd-parity bound state $E_{-,o}$ (b). The vertical dotted lines mark the charger at $x_C=0$ and the QB at $x_B=2$ and $x_B=5$, respectively.} \label{fig:photon-profiles}
\end{figure}

The profile in Fig. \ref{fig:photon-profiles} provides the real-space counterpart of the Green-function picture: each bound state is localized near the two coupling sites, yet its exponentially decaying tails retain a finite overlap across the separation, and this overlap is the origin of both the nonlocal Green-function term $G_d(E)$ and the even-odd splitting in Fig. \ref{fig:bound-spectrum}. Here, the eigenstates are obtained by exactly diagonalizing the single-excitation Hamiltonian on an open chain of $N= 281$ sites.

The squared norm of the photonic envelope in Eq. \eqref{eq:phi-real} is
%%%%%%%%%%%%%%%%%%%%%%%%%%%
\begin{equation} \label{eq:Sdef}
\begin{split}
 \mathcal S_s = {}& \frac{1}{2}\sum_x\Big(e^{-|x|/\lambda_{-,s}}\pm e^{-|x-d|/\lambda_{-,s}}\Big)^2 \\
              = {}& \coth(1/\lambda_{-,s})(1\pm e^{-d/\lambda_{-,s}})\pm d\,e^{-d/\lambda_{-,s}},
\end{split}
\end{equation}
%%%%%%%%%%%%%%%%%%%%%%%%%%%
and the full bound state is normalized by
%%%%%%%%%%%%%%%%%%%%%%%%%%%
\begin{equation}
 \mathcal{N}_s = \sqrt{1+\frac{g^2\mathcal S_s}{4J^2\sinh^2(1/\lambda_{-,s})}},
\label{eq:Norm}
\end{equation}
%%%%%%%%%%%%%%%%%%%%%%%%%%%
which sets the relative atomic and photonic fractions entering the charging analysis of Sec. \ref{sec:charging}.

\section{Bound-state beating and remote charging} \label{sec:charging}
%%%%%%%%%%%%%%%%%%%%%%%%%%%%%%%%%%%%%%%%%%%%%%%%%%%%%%%%%%%%%%%%%%%%%
In the lower-band-gap charging regime, the excitation of the charger is projected mainly onto the two lower-band dressed bound states $|\phi_{-,e}\rangle$ and $|\phi_{-,o}\rangle$. We denote the combined weight of these two states by
%%%%%%%%%%%%%%%%%%%%%%%%%%%
\begin{equation} \label{eq:PLB}
 P_\mathrm{LB} = |\langle\phi_{-,e}|\psi(0)\rangle|^2 + |\langle\phi_{-,o}|\psi(0)\rangle|^2.
\end{equation}
%%%%%%%%%%%%%%%%%%%%%%%%%%%

At the working point $d=1$ and $\delta/J=-8$, $P_\mathrm{LB} \approx 0.943$. The remaining $5.7\%$ lies in the upper-bound-state and continuum sectors. The charger excitation is therefore dominated by this two-dimensional dressed subspace. Retaining the physical projection amplitudes and omitting the upper-bound-state and continuum components gives
%%%%%%%%%%%%%%%%%%%%%%%%%%%
\begin{equation} \label{eq:psi-t}
|\psi(t)\rangle \simeq \frac{1}{\sqrt2} \big(\mathcal{N}_e e^{-iE_{-,e}t} |\phi_{-,e}\rangle
                + \mathcal{N}_o e^{-iE_{-,o}t} |\phi_{-,o}\rangle \big),
\end{equation}
%%%%%%%%%%%%%%%%%%%%%%%%%%%
and this truncated state is not renormalized: its norm $(\mathcal N_e^2+\mathcal N_o^2)/2 = P_\mathrm{LB}$ equals the lower-band weight and is below unity as the residual upper-bound-state and continuum components are dropped. Since the two TLSs are identical, the even- and odd-parity bound states carry equal charger and battery populations but evolve with different energies. Their accumulating relative phase converts the initially charger excitation into the battery and back again, producing the charging oscillation.

We denote by $Z_s = \mathcal N_s^2/2$ ($s=e,o$) the population of the battery in the parity-$s$ bound state. The stored energy of it can be approximated as
%%%%%%%%%%%%%%%%%%%%%%%%%%%
\begin{equation} \label{eq:EB-t}
 E_B(t) \simeq \Omega \big\{Z_e^2+Z_o^2 - 2Z_e Z_o \cos [(E_{-,e}-E_{-,o})t] \big\}.
\end{equation}
%%%%%%%%%%%%%%%%%%%%%%%%%%%
The first maximum occurs when the relative phase reaches $\pi$, giving the charging time $t_\mathrm{ch}\simeq \pi/|E_{-,e}-E_{-,o}|$. The maximum stored energy is then
%%%%%%%%%%%%%%%%%%%%%%%%%%%
\begin{equation} \label{eq:EBmax}
 E_{B,\max} \simeq \Omega (Z_e+Z_o)^2.
\end{equation}
%%%%%%%%%%%%%%%%%%%%%%%%%%%

% For one-column wide figures use
\begin{figure}[!htbp]
\centering
\includegraphics[width=\columnwidth]{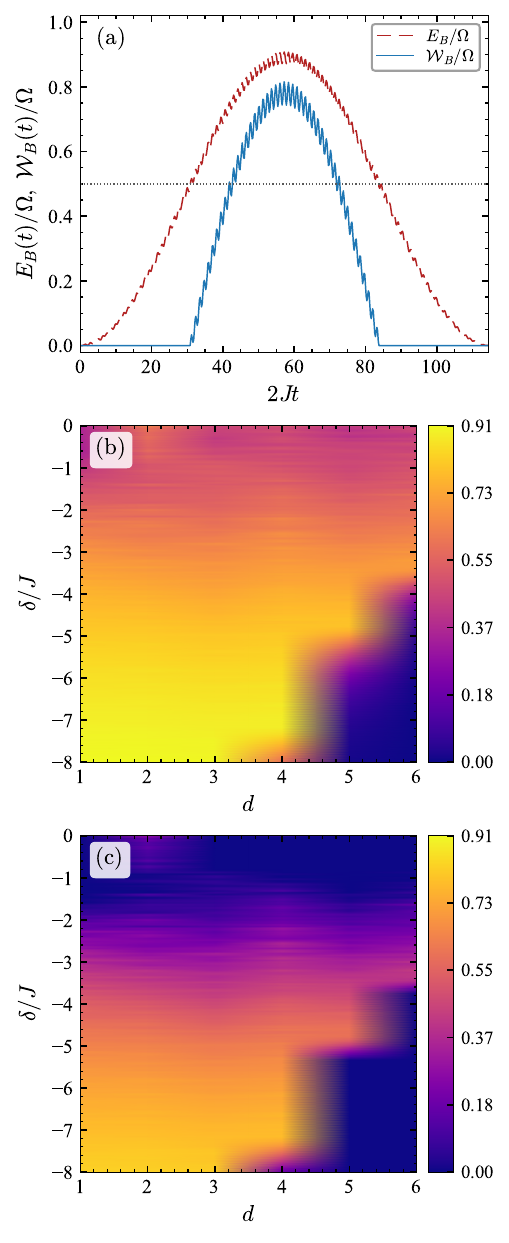}
\caption{(a) Dynamics of $E_B(t)/\Omega$ and $\mathcal{W}_B(t)/ \Omega$ for $d=1$ and $\delta/J=-8$, where the horizontal dotted line marks the threshold $E_B/\Omega=1/2$. (b) The stored maximum energy $E_{B,\max}/\Omega$ vs $d$ and $\delta/J$. (c) The charged maximum ergotropy $\mathcal{W}_{B,\max}/ \Omega$ vs $d$ and $\delta/J$. The coupling is fixed at $g/J=2$ in all the plots.} \label{fig:energy-ergotropy}
\end{figure}

For the present two-level QB, the excited-state population is $P_B(t)=\langle e_B|\rho_B(t)|e_B\rangle$. As the global state contains a single excitation, the charger-cavity states correlated with $|e_B\rangle$ and $|g_B\rangle$ belong to the orthogonal excitation sectors, hence the battery state is diagonal in the energy basis, i.e.,
%%%%%%%%%%%%%%%%%%%%%%%%%%%
\begin{equation}
 \rho_B(t) = P_B(t)|e_B\rangle\langle e_B| + [1-P_B(t)]|g_B\rangle\langle g_B|,
\end{equation}
%%%%%%%%%%%%%%%%%%%%%%%%%%%
then $E_B(t) = \Omega P_B(t)$ and $\mathcal{W}_B(t) = \Omega\max\{0,2P_B(t)-1\}$, so there is extractable work in the QB when $P_B(t) > 1/2$. At the first charging maximum, $P_B=(Z_e+Z_o)^2$, giving
%%%%%%%%%%%%%%%%%%%%%%%%%%%
\begin{equation} \label{eq:WBmax}
 \mathcal{W}_{B,\max} \simeq \Omega \max \{0, 2(Z_e+Z_o)^2-1 \}.
\end{equation}
%%%%%%%%%%%%%%%%%%%%%%%%%%%

At the working point $d=1$, $J/2\pi = 50 \, \mathrm{MHz}$, $\delta/J = -8$, and $g/J=2$, the bound-state populations are $Z_e \approx 0.466$ and $Z_o \approx 0.478$, the charging time $t_\mathrm{ch} \approx 91.42 \, \mathrm{ns}$, and from Eqs. \eqref{eq:EBmax} and \eqref{eq:WBmax} we obtain $E_{B,\max} \approx 0.891 \Omega$ and $\mathcal{W}_{B,\max} \approx 0.782 \Omega$. The exact diagonalization result in Fig. \ref{fig:energy-ergotropy}(a) gives $t_\mathrm{ch}\approx 91.06 \, \mathrm{ns}$, $E_{B,\max} \approx 0.908 \Omega$, and $\mathcal{W}_{B,\max} \approx0.816 \Omega$. The very small difference between these two sets of results arises from the upper-bound-state and continuum components omitted in the approximation \eqref{eq:psi-t}.

To assess the charging performance, we exactly diagonalize and time-evolve the charger-battery-cavity Hamiltonian in the single-excitation sector, scanning $d$ and $\delta/J$ at fixed $g/J=2$. The corresponding results are shown in Figs. \ref{fig:energy-ergotropy}(b) and \ref{fig:energy-ergotropy}(c). Over the scanned parameter range, reducing the separation of the two TLSs preserves a sizable even-odd splitting and shortens the charging time, whereas moving deeper into the lower gap increases the total TLS population of the associated bound states. These two effects together raise both the maximum battery population and the extractable work.

The propagation band and the gap support different charging dynamics. For the parameter range in Fig. \ref{fig:energy-ergotropy}, the in-band excitation spreads over the extended cavity modes, while the lower-gap bound states support coherent energy transfer from the charger to the battery. Inside the band, the charger excitation is distributed among the extended cavity modes and does not concentrate into a sufficiently inverted local battery state, thus $P_B(t)$ stays below the threshold $1/2$ and $\mathcal{W}_B(t)$ remains nearly zero within the numerical resolution. Outside the band, the same cavity supports localized dressed modes whose coherent superposition can raise $P_B(t)$ above the threshold $1/2$. The nonzero ergotropy can therefore serve as a signature of bound-state-mediated charging in the parameter range studied here.

\section{Effects of relaxation and photon loss on remote charging}
%%%%%%%%%%%%%%%%%%%%%%%%%%%%%%%%%%%%%%%%%%%%%%%%%%%%%%%%%%%%%%%%%%%%%
The out-of-band bound states inherit dissipation from both the TLSs and photonic components. We consider independent zero-temperature Markovian relaxation of the two TLSs and uniform photon loss in the array, while neglecting the pure dephasing, which would damp the even-odd beating. The density operator evolves as
%%%%%%%%%%%%%%%%%%%%%%%%%%%
\begin{equation} \label{eq:lindblad}
 \dot{\rho} = -i[H_\mathrm{rot}, \rho]
              + \gamma_C \mathcal{D}[\sigma_C^-]\rho
              + \gamma_B \mathcal{D}[\sigma_B^-]\rho
              + \kappa \sum_x \mathcal{D}[a_x]\rho,
\end{equation}
%%%%%%%%%%%%%%%%%%%%%%%%%%%
where $H_\mathrm{rot}$ is given by Eq. \eqref{eq:H-k}, $\gamma_C$ and $\gamma_B$ are the relaxation rates of the charger and the QB, respectively, $\kappa$ is the uniform photon-loss rate of the array, and $\mathcal{D}[O]$ is the Lindblad dissipator with $\mathcal{D}[O]\rho = O\rho O^\dagger - \{O^\dagger O, \rho\}/2$.

% For one-column wide figures use
\begin{figure}[!htbp]
\centering
\includegraphics[width=\columnwidth]{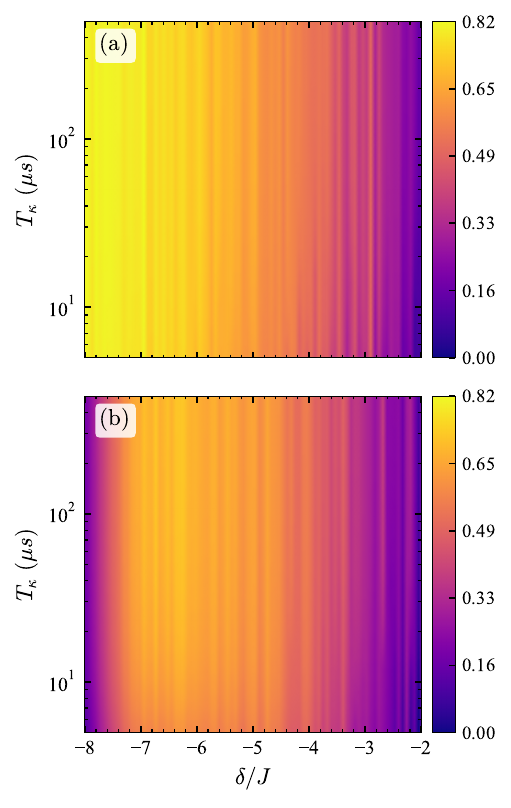}
\caption{The maximum ergotropy $\mathcal{W}_{B,\max}/\Omega$ obtained from the Lindblad dynamics \eqref{eq:lindblad} for (a) $d=1$ and (b) $d=3$. Other parameters are $g/J=2$, $T_1^C=30 \mu\mathrm{s}$, and $T_1^B=200 \mu\mathrm{s}$. The cavity lifetime $T_\kappa$ varies from $5 \,\mu\mathrm{s}$ to $500\,\mu\mathrm{s}$ on a logarithmic scale.} \label{fig:open-system}
\end{figure}

In this study, we take $J/2\pi = 50 \, \mathrm{MHz}$ and $\omega_c/2\pi = 6 \, \mathrm{GHz}$, representative of the superconducting coupled-resonator array \cite{Mirhosseini2018,Anderson2016}. Moreover, we consider $T_1^C= 30\,\mu\mathrm{s}$, $T_1^B= 200\,\mu\mathrm{s}$ (i.e., we take $\gamma_C>\gamma_B$ to represent a charger with a shorter relaxation time than that of the battery), and scanning $T_\kappa=\kappa^{-1}$ from $5\,\mu\mathrm{s}$ to $500\,\mu\mathrm{s}$. Based on this setup, we calculate the stored energy and charged ergotropy of the QB directly from Eq. \eqref{eq:lindblad}. Figure \ref{fig:open-system} shows the maximum ergotropy as a function of $\delta/J$ and $T_\kappa$ for $d=1$ and $d=3$, with fixed $g/J=2$. At $\delta/J=-8$ and $T_\kappa=500\,\mu\mathrm{s}$, increasing the separation from $d=1$ to $d=3$ reduces $\mathcal{W}_{B,\max}/\Omega$ from the value of about $0.81$ to $0.17$. The larger separation $d$ weakens the even-odd splitting and prolongs the time required to reach the first charging maximum. Consequently, the dissipation acts in a longer time interval and the charging performance is degraded. At fixed separation, moving deeper into the lower gap increases the total TLS population of the relevant bound states and reduces their sensitivity to cavity photon loss. Increasing $T_\kappa$ therefore has a strong effect on the charged ergotropy in the vicinity of the band edge, where the bound states contain a large photonic fraction.

At the deep-gap working point, the charging time is much shorter than the TLS relaxation times and the total TLS population of each relevant bound state exceeds $90\%$. The first charging maximum is therefore only weakly affected by the photon loss considered here. The localization length acts as a common control factor for the interaction range and the sensitivity to cavity decay: the states closer to the band edge extend over more cavity sites and mediate coupling of the TLSs over a larger separation, whereas the deeper bound states suppress the cavity photon loss. On the experimental side, the required hopping and coupling scales considered in Fig. \ref{fig:open-system} are compatible with the superconducting coupled-resonator arrays and the high-coherence transmons \cite{Place2021}.

\section{Conclusion}
%%%%%%%%%%%%%%%%%%%%%%%%%%%%%%%%%%%%%%%%%%%%%%%%%%%%%%%%%%%%%%%%%%%%%
In conclusion, we have considered a charger-battery model consisting of two TLSs locally coupled to the different sites of a one-dimensional coupled cavity array. Our results revealed that the finite photonic band can serve not only as a barrier to radiative escape but also as a coherent channel for remote charging. Outside the propagation band, the charger, QB, and cavity array support atom-photon bound states, and the charging dynamics is governed predominantly by the two lower-band parity-resolved bound states. The overlap of their photonic components produces an even-odd energy splitting that drives reversible energy exchange between the charger and QB. Such a relation between spectrum and system dynamics relates the bound-state energy and TLS populations directly to the charging time, stored energy, and charged ergotropy.

The localization of the bound states provides a controllable way for balancing the interaction range of the TLSs and dissipation of the total system. Physically, a deep detuning reduces the photonic contribution and protects the first charging maximum, whereas the states closer to the band edge extend the interaction over a larger separation, and in the circuit-QED arrays, for example, the balance can be adjusted through tuning the qubit frequency, the interresonator hopping, and the local light-matter coupling \cite{QED1,QED2}. Moreover, local frequency control or a switchable coupling can suppress the energy backflow to the charger after the first charging maximum. An interesting direction for future study is to extend the similar mechanism to multiple energy storage units and engineered photonic bands, which can enable remotely addressable charger-battery networks.

\section*{Acknowledgments}
This work was supported by the National Natural Science Foundation of China (Grants No. 12275212 and No. 12575027), the Youth Innovation Team of Shaanxi Universities (Grant No. 24JP177), and the Natural Science Foundation of Shaanxi Province (Grant No. 2025JC-YBQN-055).

\section*{DATA AVAILABILITY}
The data that support the findings of this article are openly available \cite{data}.

%BibTeX users please use
%\bibliographystyle{}
%\bibliography{}

\begin{thebibliography}{99}
\bibitem{data}
Data will be made available on Zenodo for the final published version.
\end{thebibliography}
%\Non-BibTeX users please use

\end{document}